\documentclass[12pt]{iopart}
\usepackage{graphicx}
\usepackage{dcolumn}
\usepackage{bm}

\begin{document}
\title{Quantum Theory of Fiber Bragg Grating Solitons}
\author{Ray-Kuang Lee\dag\ddag and  Yinchieh Lai\dag
\footnote[3]{To whom correspondence should be addressed (yclai@mail.nctu.edu.tw)}}
\address{\dag Institute of Electro-Optical Engineering, National Chiao-Tung University, Hsinchu, Taiwan, Republic of China\\ \ddag National Center for High-Performance Computing, Hsinchu, Taiwan, Republic of China}
\date{\today}

\begin{abstract}
Following the pioneering work of Prof. Hermann A. Haus, a general quantum theory for bi-directional nonlinear optical pulse propagation problems is developed and applied to study the quantum properties of fiber Bragg grating solitons.
Fiber Bragg grating solitons are found to be automatically amplitude squeezed after passing through the grating and the squeezing ratio saturates after a certain grating length. The optimal squeezing ratio occurs when the pulse energy is slightly above the fundamental soliton energy. One can also compress the soliton pulsewidth and enhance the squeezing simultaneously by using an apodized grating, as long as the solitons evolve adiabatically.
\end{abstract}

\pacs{42.50.Dv, 42.81.Dp, 42.70.Qs} 
\submitto{\JOB\\Special Issue on Fluctuations and Noise in Photonics and Quantum Optics}
\maketitle

\section{Introduction}
Solitons are particle-like waves that propagate in dispersive or absorptive media without changing their pulse shapes and can survive after collisions.
Various types of optical soliton phenomena have been studied extensively in the area of nonlinear optical physics.
These include the nonlinear Schr{\" o}dinger solitons in dispersive optical fibers, spatial and vortex solitons in photorefractive materials/waveguides, and cavity solitons in resonators \cite{Kivshar01}.
Solitons in optical fibers were first predicted by Hasegawa and Tappert \cite{Hasegawa73} in 1973 and were first observed experimentally by Mollenauer et al. \cite{Mollenauer80} in 1980.
Since then, the idea of using solitons for long-haul optical transmission has been an attractive research area with rapid progress \cite{Iannone98}.
In 1986, Gordon and Haus pointed out that the spontaneous emission from optical amplifiers in the optical link can cause the timing jitter variance of the solitons to be proportional to the cubic of the propagation distance \cite{Gordon86}.
This effect, known as the Gordon-Haus effect, will place an upper limit on the achievable bit rate in the long-haul soliton communication systems, even though it was found out later that one can somehow overcome this limitation by using guiding filters \cite{Mecozzi91, Kodama92}.

The spontaneous emission effects on soliton propagation already calls for a rigorous quantum theory for soliton propagation, even though it may still be possible to treat them semi-classically. In 1987, Carter and Drummond et al. \cite{Carter87, Drummond87} used the positive-$P$ representation approach to transform the quantum nonlinear Schr{\" o}dinger equation into stochastic nonlinear equations with noise terms.
They then solved the stochastic nonlinear equations numerically and showed that solitons are quantum mechanically squeezed during propagation.
The squeezing ratio of the quantum solitons was later calculated with the inclusion of the homodyne detection scheme \cite{Carter89}.
Since then, the quantum theory of traveling-wave optical solitons has been intensively developed and several approaches have been successfully carried out. The studied soliton systems include the family of nonlinear Schr{\" o}dinger solitons \cite{Drummond87, Lai89a, Lai89b, HHaus90, Schmidt00} as well as the self-induced-transparency solitons \cite{Lai90}.
The quantum theory of nonlinear Schr{\" o}dinger solitons has also been directly applied to study the soliton communication problems \cite{Haus91, Lai93a, Lai93b}.

The quantum theory of nonlinear Schr{\" o}dinger solitons turns out to be an very interesting one.
It was sometimes less known that the quantum nonlinear Schr{\" o}dinger equation is the evolution equation of one-dimensional bosonic system with $\delta$-function interaction under the second quantization framework \cite{Huang87}.
By solving the problem in the Schr{\" o}dinger picture using the Bethe's ansatz method, one can construct bound-state eigensolutions which are closely related to the soliton phenomenon.
In 1989, Lai and Haus noticed this correspondence and constructed the exact soliton states by the Bethe's ansatz method \cite{Lai89b}.
At the same time they also developed an approximate nonlinear analysis based on the time-dependent Hartree approximation \cite{Lai89a}.
In the subsequent year, a quantum soliton perturbation theory based on the linearization approach was successfully carried out by Haus and Lai \cite{HHaus90}.
The squeezing ratio of the soliton parts can be predicted by a simple analytic formulation and the optimal local oscillator pulse shape can be determined.
This quantum soliton perturbation theory was later re-formulated to make it more clear \cite{Lai93a} and eventually led to the development of the {\it back-propagation method} \cite{YLai95}, which is a general numerical method for calculating the quantum noises of nonlinear pulse propagation problems based on the linearization approximation.
The aim of the present paper is to further pursue the theoretical development along this line by generalizing the theory of quantum solitons to bi-directional nonlinear pulse propagation problems and to study the quantum properties of Bragg grating solitons in particular.

On the side of experimental progresses, many experiments for actually measuring the soliton squeezing have been carried out since 1990, with the advance of stable pulse lasers and high quantum efficiency detectors at the optical communication wavelengths. 
To provide a stable experimental setup against the environmental fluctuations, Shirasaki and Haus proposed to use a Sagnac loop in soliton squeezing experiments for measuring the quadrature squeezing \cite{Shirasaki}.
The scheme is actually a squeezed vacuum state generator that can output squeezed vacuum states. 
Soliton squeezing from a Sagnac fiber interferometer was first observed by Rosenbluh and Shelby in 1991 (1.7 dB below the shot noise)\cite{Rosenbluh}.
Using the same setup but at the different wavelength, Bergman and Haus also succeeded to observe squeezing with non-soliton pulses (4.9 dB below the shot noise)\cite{Bergman91} in the same year.
Since then, larger quadrature squeezing from fibers has been obtained with a gigahertz Erbium-doped fiber lasers to suppress the guided acoustic-wave Brillouin scattering (GAWBS) and a 6.1 dB noise reduction below the shot noise has been reported \cite{Yu01}.
One can also produce amplitude squeezing by using solitons with an external spectral filter \cite{Friberg} or with an imbalance Sagnac loop interferometer \cite{Krylov98}.
Squeezing generation with an imbalance Sagnac loop interferometer in the normal-dispersion region has also been demonstrated \cite{Krylov99}.
Recently, quantum correlation between different spectral components of the squeezed soliton has also been studied extensively in the literature \cite{Spalter}. 
The above brief description summarizes the exciting development that has been carried out for squeezing generation using optical fibers. 

In recent years, with the advance of new fabrication technologies, it becomes more feasible to actually utilize one- or higher dimensional periodic dielectric structures (or especially the photonic bandgap crystals) \cite{Joannopoulos95} for modifying the properties of the spontaneous emission as well as the propagation of waves.
The fiber Bragg gratings (FBGs) are one-dimensional periodic structures with weak index modulation.
The simplest way to induce 1-D Bragg gratings inside optical fibers is with the side-illumination of the UV interference lights.
The FBG formed this way can be viewed as a 1-D photonic bandgap crystal for the guiding mode of the single-mode fiber \cite{Erdogan97}. 
Interesting soliton phenomena known as the fiber Bragg grating solitons \cite{AAceves89, BEggleton96} can be found if the fiber Bragg grating has the third-order Kerr nonlinearity.
The propagation of optical pulses inside the nonlinear FBG can be described by the nonlinear coupled mode equations (NCMEs).
Intuitively the fiber Bragg grating soliton is formed when the input pulse has the suitable pulse-width and peak intensity such that the nonlinear Kerr effect is large enough to compensate the high anomalous dispersion near one of the bandedges of the FBG.

From the theoretical point of view, fiber Bragg grating solitons belong to the class of bi-directional pulse propagation problems, where the quantum theory is still lack of enough consideration.
Most of the previous studies on fiber Bragg grating solitons have been on the classical effects and there is almost no reported result on their quantum properties.
It is the aim of this study to bridge this gap by developing a general quantum theory for bi-directional pulse propagation problems and particularly to apply the theory for studying the case of fiber Bragg grating solitons.
It will be shown that the fiber Bragg grating soliton pulses will quantum-mechanically get amplitude-squeezed after passing through the fiber grating and the squeezing ratio will be calculated theoretically.
We use the linearization approach to study the quantum effects of FBG solitons by extending the {\it back-propagation method} previously developed \cite{YLai95} to the case of nonlinear bi-directional propagation problems.
By following the same spirit of the back-propagation method, we will first derive a set of linear adjoint equations from the linearized nonlinear coupled mode equations in such a way that any inner product between the solutions of the two equation sets is conserved during the time evolution.
In this way, the variance of the measured operator as well as its squeezing ratio can be calculated readily for a given measurement characteristic function.
The squeezing ratio of FBG solitons will be found to exhibit interesting relations with the fiber grating length as well as with the intensity of the input pulse.

In contrast to uniform fiber gratings, nonuniform fiber gratings with chirp and/or apodization have shown some potentials for pulse compression applications \cite{Eggleton00}.
The theory and experiment for Bragg grating solitons propagated in apodized FBGs have also been developed and demonstrated \cite{Lenz98, Eggleton99}.
Some new results on slowing down FBG solitons in apodized fiber gratings have also been carried out \cite{Mak03}.
In the present paper the quantum fluctuations of FBG solitons in nonuniform fiber gratings will also be studied.
 We find that one can compress the soliton pulsewidth and enhance the squeezing simultaneously by using an apodized grating, as long as the solitons evolve adiabatically.

The paper is organized as follows: in section \ref{secBGS} the model of the nonlinear coupled mode equations for bi-directional wave propagation in a uniform FBG is reviewed.
In section \ref{secQBGS} we use the {\it back propagation method} to calculate the quantum fluctuations of FBG solitons and the characteristics of the squeezing ratio are discussed.
In section \ref{secAPD} the quantum fluctuations of FBG solitons in nonuniform fiber Bragg gratings are studied.
 Finally, a brief conclusion is given in section \ref{secCon}.

\section{Fiber Bragg Grating Solitons}
\label{secBGS}
\begin{figure} 
\begin{center}
\includegraphics[height=3.0in]{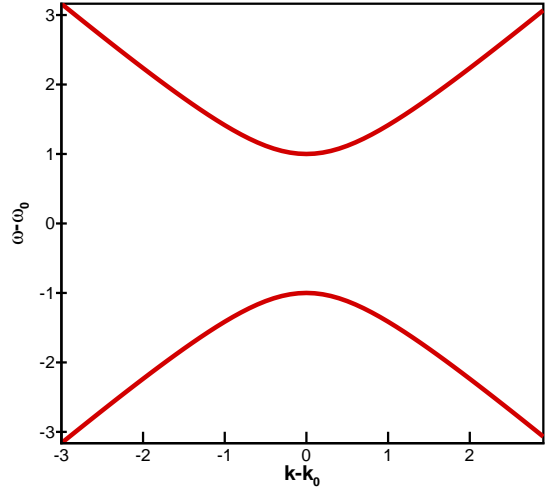}
\caption{Dispersion relation of linear fiber Bragg gratings.}
\label{figgap}
\end{center}
\end{figure}

We start the derivation from considering the linear wave propagation problem in a one dimensional periodic structure:
\begin{eqnarray}
\frac{\partial^2 E}{\partial z^2}-\frac{n^2(z)}{c^2}\frac{\partial^2 E}{\partial t^2}=0
\end{eqnarray}
Here the dielectric constant $n^2(z) = \bar{n}^2+\tilde{\epsilon}(z)$ is a periodic function of the propagation distance, with the spatial average refractive index $\bar{n}$ and the period $\Lambda$.
We are interested in the light field at the frequency near the Bragg condition $\omega_0 = k_0 c/\bar{n}$, where $c$ is the light velocity in free space, and $k_0 = \pi/\Lambda$.
For FBGs, we can expand the periodic index perturbation function $\tilde{\epsilon}(z)$ by the Fourier series and only keep the phase-matching $\pm 1$ order terms, i.e. $\tilde{\epsilon}(z) = 2 \tilde{\epsilon}_0 \cos (2 k_0 z)$.
One then decomposes the light field into the forward ($U_a$) and backward ($U_b$) propagation pulses, $E(z,t) = U_a(z,t)e^{-i(\omega t -k_0 z)}+U_b(z,t)e^{-i(\omega t +k_0 z)}+c.c.$, and obtain the following linear coupled mode equations:
\begin{eqnarray}
\frac{1}{v_g}\frac{\partial}{\partial t} U_a(z,t)+\frac{\partial}{\partial z} U_a = i\delta U_a+i\kappa U_b\\
\frac{1}{v_g}\frac{\partial}{\partial t} U_b(z,t)-\frac{\partial}{\partial z} U_b = i\delta U_b+i\kappa U_a
\end{eqnarray}
where $v_g = \bar{n}/c$ is the group velocity of the pulses,
$\delta = \omega-\omega_0$ is the wavelength detuning parameter,
and $\kappa = \omega_0\tilde{\epsilon}/2\bar{n} c$ is the coupling coefficient.
In Fig. \ref{figgap} one can see that the dispersion relation of this set of coupled-mode equations has a bandgap at the frequency (wavelength) that satisfies the Bragg condition $k = k_0$.

If the third order nonlinearity of the optical fiber needs to be taken into account, one can model the problem by using the above coupled mode equations that describe the coupling between the forward and the backward propagating waves in a uniform FBG with the addition of the self-phase and cross-phase modulation effects.
With the nonlinear terms, this set of NCMEs has analytical soliton solutions for the case of infinite grating length, as is shown by Aceves and Wabnitz with the introduction of the massive Thirring model \cite{AAceves89}.
However, for nonlinear FBGs of finite length, no analytic solution can be found.
So in our studies we directly use the finite difference numerical simulation method with the parameters based on the first experiment reported in the literature \cite{BEggleton96}. 
We consider a 60 $ps$ full width at half maximum (FWHM) {\it sech}-shaped pulse incidents into a uniform grating with 15.0 $cm^{-1}$ wavenumber detuning from the center of the bandgap.
The coupling strength of the fiber grating is 10 $cm^{-1}$, the nonlinear coefficient $\Gamma$ is 0.018 $cm/GW$, and the group velocity $v_g$ is chosen to be $c/n$ with $n=1.5$ and $c$ being the speed of light in free space.
When the input peak intensity is below the required value for forming a solitary pulse in the FBG (about 4.5 $GW/cm^2$ in this case), the peak intensity of the pulse will decrease along the propagation.
On the other hand, when the input peak intensity is above 4.5 $GW/cm^2$, the peak intensity of the pulse oscillates during the propagation within the grating.
Only when the nonlinearity can exactly compensate the dispersion induced by the FBG, one can have a stable solitary pulse inside the grating.

\section{Quantum Fiber Bragg Grating Solitons}
\label{secQBGS}
After finding these classical solutions, we now turn to the calculation of the quantum properties.
The Hamiltonian of the system can be expressed by the forward ($\hat{U_a}$) and backward ($\hat{U_b}$) field operators as follows \cite{ZCheng96}:
\begin{eqnarray}
\label{eqHam}
{\cal H} = &-& v_g \{\delta \int dz\, (\hat{U}_a^\dag\hat{U}_a + \hat{U}_b^\dag\hat{U}_b)+ i\int dz\,(\hat{U}_a^\dag\frac{\partial}{\partial z}\hat{U}_a -\hat{U}_b^\dag\frac{\partial}{\partial z}\hat{U}_b)\\\nonumber
&+&\kappa \int dz\, (\hat{U}_a^\dag \hat{U}_b + \hat{U}_b^\dag \hat{U}_a)\\\nonumber
&+&\frac{\Gamma}{2}\int dz\, (\hat{U}_a^\dag\hat{U}_a^\dag\hat{U}_a\hat{U}_a+\hat{U}_b^\dag\hat{U}_b^\dag\hat{U}_b\hat{U}_b)\\\nonumber
&+&\Gamma\int dz\,(\hat{U}_a^\dag\hat{U}_b^\dag\hat{U}_b\hat{U}_a+\hat{U}_b^\dag\hat{U}_a^\dag\hat{U}_a\hat{U}_b)\}
\end{eqnarray}
Here the field operators $\hat{U}_a$ and $\hat{U}_b$ must obey the usual equal time Bosonic commutation relations.
\begin{eqnarray*}
&&[\hat{U}_a(z_1, t),\hat{U}_a^\dag(z_2, t)] = \delta(z_1-z_2)\\
&&{[}\hat{U}_b(z_1, t),\hat{U}_b^\dag(z_2, t)] = \delta(z_1-z_2)\\
&&{[}\hat{U}_a(z_1, t),\hat{U}_a(z_2, t)] = {[}\hat{U}^\dag_a(z_1, t),\hat{U}_a^\dag(z_2, t)] = 0\\
&&{[}\hat{U}_b(z_1, t),\hat{U}_b(z_2, t)] = {[}\hat{U}^\dag_b(z_1, t),\hat{U}_b^\dag(z_2, t)] = 0\\
&&{[}\hat{U}_a(z_1, t),\hat{U}_b(z_2, t)] = {[}\hat{U}_a(z_1, t),\hat{U}_b^\dag(z_2, t)] =0
\end{eqnarray*}

\begin{figure}
\begin{center}
\includegraphics[width=3.0in]{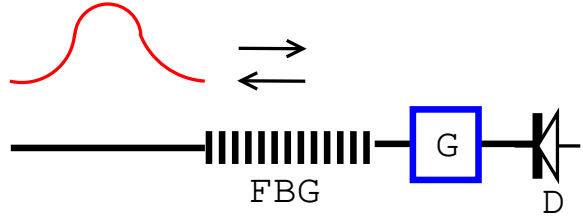} 
\caption{Measurement scheme of direct detection for observing FBG soliton squeezing. Here $G$ is a gating device which will block out all the transmittive pulses but the first one; $D$ is an optical detector.}
\label{msr1}
\end{center}
\end{figure}

In the Heisenberg picture we can derive the quantum nonlinear coupled mode equations (QNCMEs) by the Hamiltonian in Eq. (\ref{eqHam}):
\begin{eqnarray}
\label{QNCME1}
\frac{1}{v_g}\frac{\partial}{\partial t} \hat{U}_a(z,t)+ \frac{\partial}{\partial z} \hat{U}_a &=& i\delta \hat{U}_a+i\kappa\hat{U}_b+i\Gamma \hat{U}_a^\dag \hat{U}_a \hat{U}_a+2 i\Gamma \hat{U}_b^\dag \hat{U}_b \hat{U}_a\\
\label{QNCME2}
\frac{1}{v_g}\frac{\partial}{\partial t} \hat{U}_b(z,t)- \frac{\partial}{\partial z} \hat{U}_b &=& i\delta \hat{U}_b+i\kappa \hat{U}_a+i\Gamma \hat{U}_b^\dag \hat{U}_b \hat{U}_b+2 i\Gamma \hat{U}_a^\dag \hat{U}_a \hat{U}_b
\end{eqnarray}
This derivation automatically proves that the QNCMEs preserve the commutation brackets.
The form of QNCMEs in Eqs. (\ref{QNCME1}-\ref{QNCME2}) is the same as the classical NCMEs, except that it is now a set of quantum operator equations.
For optical solitons with a large average photon number, we can safely use the linearization approximation to study their quantum effects.
By setting
\begin{eqnarray*}
\hat{U}_a(z,t) &=& U_{a0}(z,t) +\hat{u}_a(z,t)\\
\hat{U}_b(z,t) &=& U_{b0}(z,t) +\hat{u}_b(z,t)
\end{eqnarray*}
where $U_{a0}$ and $U_{b0}$ are classical fiber Bragg grating soliton solutions of NCMEs, one can obtain a set of linear quantum operator equations that describe the evolution of the quantum fluctuations associated with the fiber Bragg grating solitons after the linearization approximation is performed.
\begin{eqnarray}
\label{ptbeq}
\hspace{-1.0in}\frac{1}{v_g}\frac{\partial}{\partial t} \left(\begin{array}{c}\hat{u}_a\\\hat{u}_b\end{array}\right) &=& \left(\begin{array}{cc} -\frac{\partial}{\partial z}+i\delta+2 i\Gamma|U_{a0}|^2+2 i\Gamma|U_{b0}|^2
& i\kappa+2 i\Gamma U_{a0}U_{b0}^\ast\\
i\kappa+2 i\Gamma U_{a0}^\ast U_{b0} & \frac{\partial}{\partial z}+i\delta+2 i\Gamma|U_{a0}|^2+2 i\Gamma|U_{b0}|^2\end{array}\right)\left(\begin{array}{c}\hat{u}_a \\ \hat{u}_b \end{array}\right)\nonumber\\
\hspace{-1.0in}&+& \left(\begin{array}{cc} i\Gamma U_{a0}^2 & 2 i\Gamma U_{a0} U_{b0}\\
2i\Gamma U_{a0}U_{b0} & i\Gamma U_{b0}^2 \end{array}\right)\left(\begin{array}{c}\hat{u}^\dag_a \\ \hat{u}^\dag_b \end{array}\right)\\\nonumber
\end{eqnarray}
This set of equations is still a set of quantum operator equations for the quantum perturbation field operators $\hat{u}_a(z,t)$ and $\hat{u}_b(z,t)$, which have to satisfy the same equal time commutation relations as the unperturbed field operators $\hat{U}_a(z,t)$ and $\hat{U}_b(z,t)$.

To solve the linear quantum operator equations in Eq. (\ref{ptbeq}) by the back-propagation method, we define the inner product operation according to 
\begin{eqnarray}
\label{eqinner}
\langle\vec{f}|\vec{\hat{g}}\rangle &=& \frac{1}{2}\int dz\,[f^\ast_a \hat{g}_a+f_a \hat{g}^\dag_a+f^\ast_b \hat{g}_b+f_b \hat{g}^\dag_b]
\end{eqnarray}
We then seek for a set of adjoint equations which satisfies the following property:
\begin{eqnarray}
\label{eqconserve}
\frac{d}{d\,t}\langle\vec{u}^A|\vec{\hat{u}}\rangle = 0
\end{eqnarray}
In other words, the inner product between the solutions of the two equation sets is preserved along the time axis.
With this desired property, we can express the inner product of the output quantum perturbation operator with a projection function in terms of the input quantum field operators which have the know quantum characteristics.
This will allow us to calculate the quantum fluctuations of any inner product between the output quantum operator with a given projection function.
It is not difficult to show that the set of adjoint equations for the perturbed QNCMES in Eq. (\ref{ptbeq}) is given by:
\begin{eqnarray}
\label{adjeq}
\hspace{-1.0in}\frac{1}{v_g}\frac{\partial}{\partial t}\left(\begin{array}{c}u^A_a\\u^A_b\end{array}\right) &=&
\left(\begin{array}{cc} -\frac{\partial}{\partial z}+i\delta+2 i\Gamma|U_{a0}|^2+2 i\Gamma|U_{b0}|^2
& i\kappa+2 i\Gamma U_{a0} U_{b0}^\ast\\
i\kappa+2 i\Gamma U_{a0}^\ast U_{b0} & \frac{\partial}{\partial z}+i\delta+2 i\Gamma|U_{a0}|^2+2 i\Gamma|U_{b0}|^2\end{array}\right)\left(\begin{array}{c}u^A_a \\ u^A_b \end{array}\right)\nonumber\\
\hspace{-1.0in}&+& \left(\begin{array}{cc} -i\Gamma U_{a0}^2 & -2 i\Gamma U_{a0} U_{b0}\\
-2i\Gamma U_{a0}U_{b0} & -i\Gamma U_{b0}^2 \end{array}\right)\left(\begin{array}{c} u^{A\ast}_a \\ u^{A\ast}_b \end{array}\right)
\end{eqnarray}

Actually under the linearization approximation the measurement of a physical quantity can always be expressed as an inner product between a measurement characteristic function and the perturbed quantum field operator \cite{YLai95}. By back-propagating the classical adjoint equation set with the measurement characteristic function as the initial condition, one can express the measured inner product as an inner product between the input quantum operator and the back-propagated measurement function. It is then quite easy to calculate the quantum uncertainties of the measured quantity if the quantum characteristics of the input quantum field operators are known. Typically we will assume the input quantum state to be a coherent state.
The squeezing ratio of the measured quantity is defined according to:
\begin{eqnarray}
R(T) &=& \frac{var[\langle \vec{f}(z)| \vec{\hat{u}}(z,t=T)]}{var[\langle \vec{f}(z)| \vec{\hat{u}}(z,t=0)]}\\\nonumber
&=& \frac{var[\langle \vec{F}_T(z)| \vec{\hat{u}}(z,t=0)]}{var[\langle \vec{f}(z)| \vec{\hat{u}}(z,t=0)]}
\end{eqnarray}
Here $var[\cdot]$ means the variance, $\vec{f}(z)$ is the original projection function and $\vec{F}_T(z)$ is the back-propagated projection function.
The choice of the characteristic function $\vec{f(z)}=(f_a(z), f_b(z))^T$ will depend on the measurement to be performed. When the measured quantity is the photon number of the pulse, the characteristic function $f_a(z)$ is simply the normalized classical output pulse \cite{HHaus90} and $f_b(z)=0$. If the homodyne detection is used, then $f_a(z)$ is the normalized local oscillator pulse.  

In Fig. \ref{msr1} we illustrate the possible direct detection scheme for measuring the photon number squeezing of the
FBG solitons. To avoid the complication due to the multiple transmitted pulses of the FBG, it may be necessary to use a time-gating device to make sure that only the first transmitted pulse is detected.

\begin{figure}
\begin{center}
\includegraphics[width=3.0in]{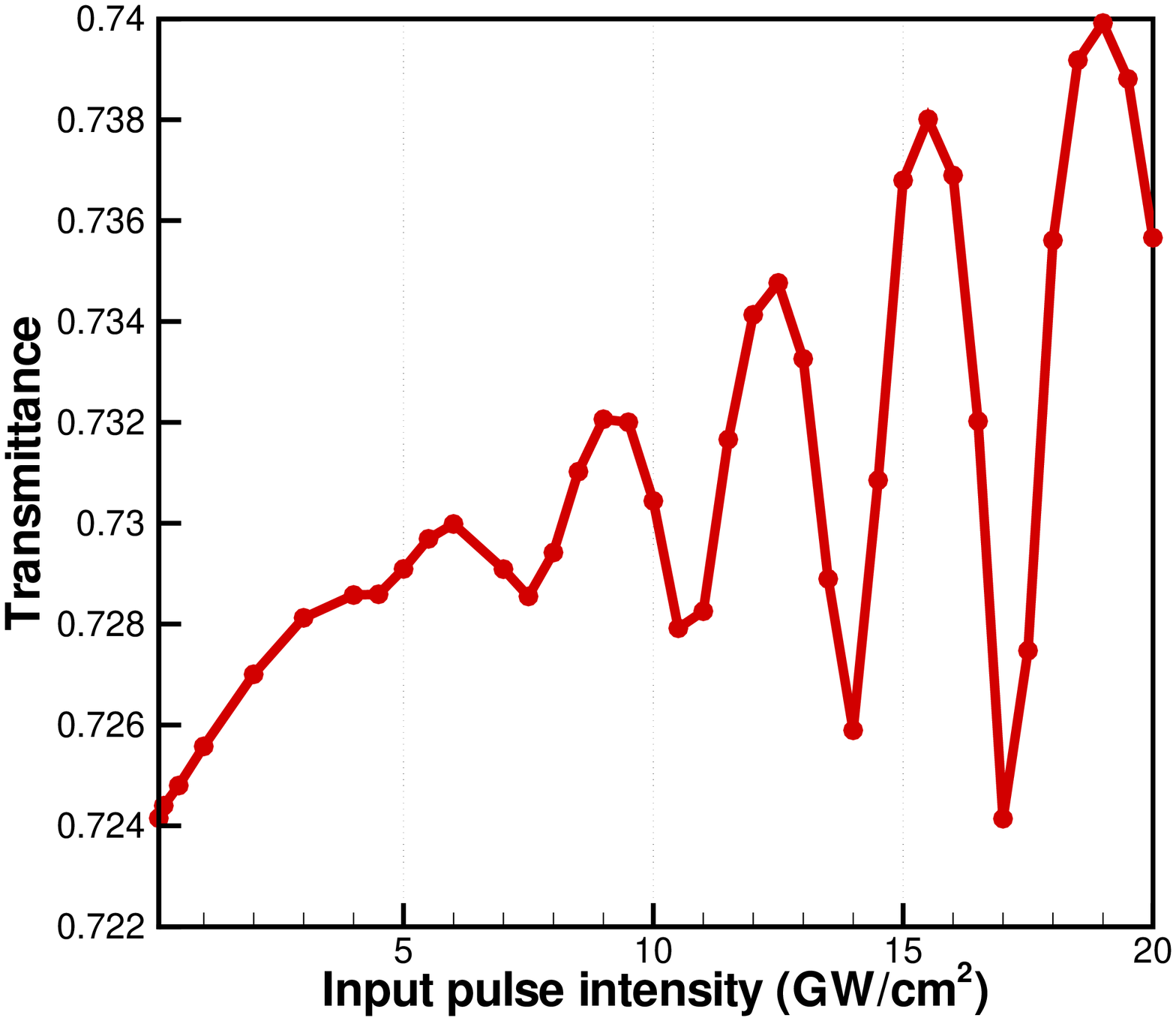} 
\includegraphics[width=3.0in]{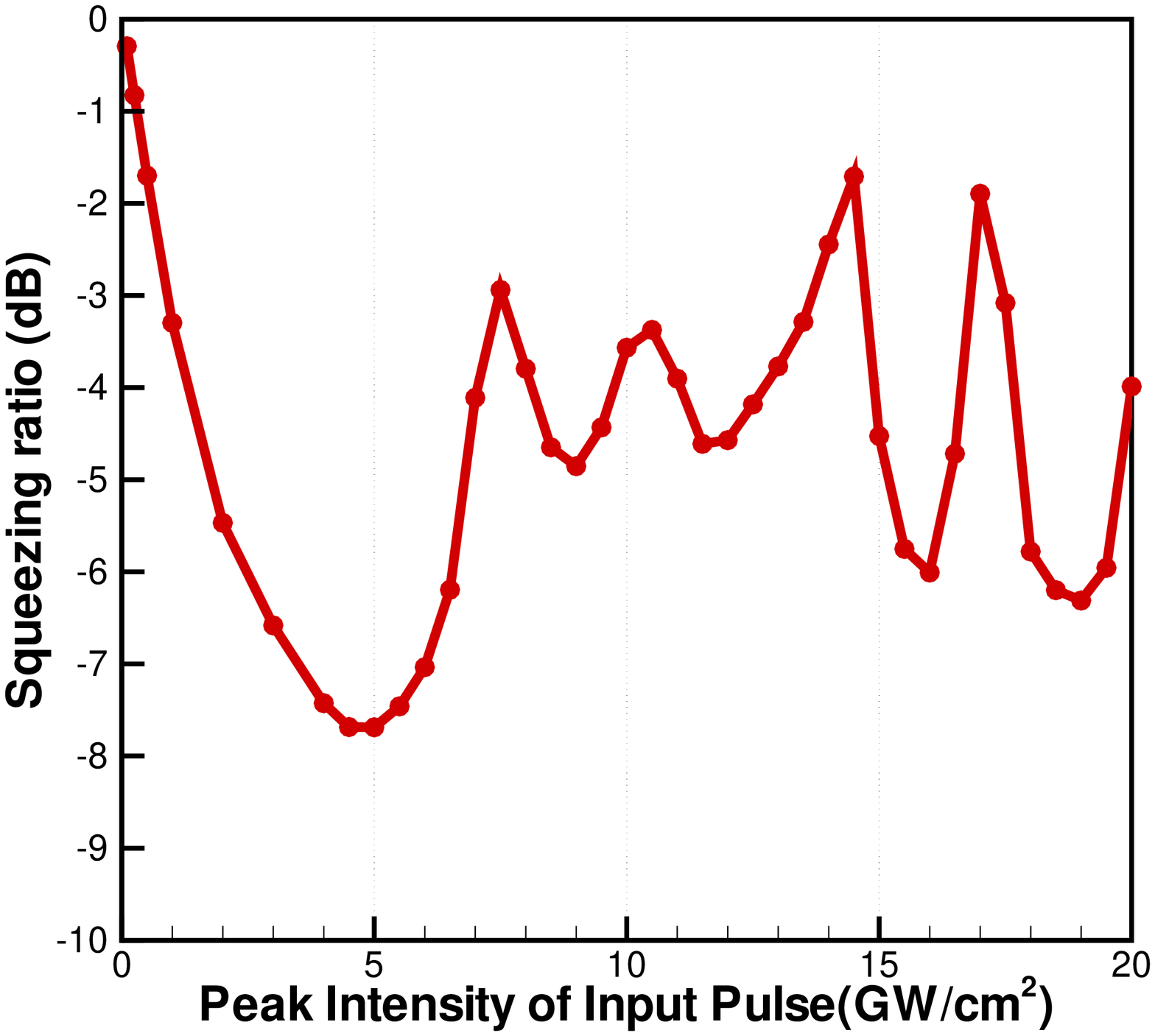} 
\caption{Transmittance (top) and photon number squeezing ratio (bottom) for fiber Bragg grating solitons with different input intensities.}
\label{SRIfig}
\end{center}
\end{figure}

Based on the formulation given above, we now apply the developed quantum theory to the case of a pulse incident into a uniform nonlinear FBG and calculate the photon number fluctuations of the first transmitted pulse.
First of all, the transmittance of a 50 $cm$ long FBG versus different intensities of the input pulse is plotted in the top curve of Fig. \ref{SRIfig}.
It can be seen that the transmittance of the FBG is a nonlinear oscillating function of the input pulse intensity. The calculated photon number squeezing ratio is shown in the bottom of Fig. \ref{SRIfig} with the same parameters.
One can see that the squeezing ratio decreases monotonically when the input intensity is below the intensity of the fundamental soliton. 
The optimum squeezing ratio occurs when the pulse energy of soliton is slightly large than that of the fundamental soliton.
The curve begins to oscillate strongly when the intensity is much larger than that of the fundamental soliton. One important observation is that the oscillation behavior of the FBG transmittance and the squeezing ratio matches very well. That is, the squeezing ratio has a local minimum when the transmission has a local maximum. This agrees with the intuitive expectation that larger amplitude squeezing should occur when the transmittance curve is saturated. Another way to intuitively understand the results is as follows. The periodic grating structure acts like a spectral filter which can filter out the noisier high frequency components in the soliton spectrum and produce a net photon number (amplitude) squeezing effect just as in the previous soliton amplitude squeezing experiments where a spectral filter is cascaded after a nonlinear fiber \cite{Friberg, TOpatrny02}.

\begin{figure}
\begin{center}
\includegraphics[width=3in]{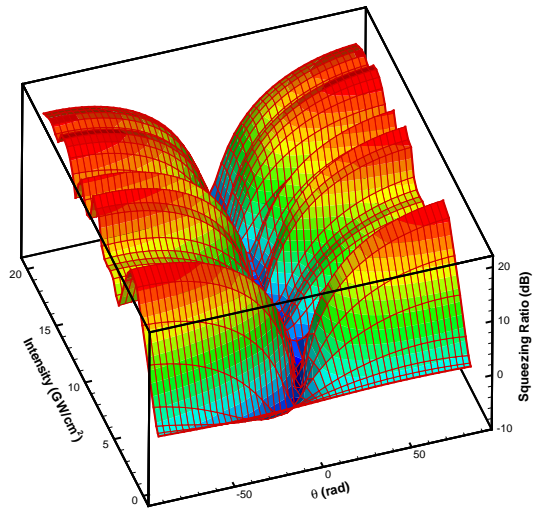}
\caption{Squeezing ratio of the quadrature field versus the input intensity and the local oscillator phase.}
\label{SSIfig}
\end{center}
\end{figure}

To make sure that the FBG solitons are actually amplitude squeezed, we perform another calculation to simulate the squeezing ratio when the homodyne detection scheme is used and assume that the local oscillator pulse is exactly the classical output pulses. With the homodyne detection scheme, one has the additional degree of freedom to adjust the relative phase between the local oscillator and the signal.
Fig. \ref{SSIfig} plots the squeezing ratio for different input intensities and for different local oscillator phases when propagating through a constant FBG length ($ 50 cm^2$). One can see that for small input intensities (below $5 GW/cm^2$) the squeezing direction is close to but not exactly in the photon number (or amplitude) quadrature, $\theta = 0$. However, when the intensity of the FBG soliton is large enough, the squeezing direction will be in the photon number quadrature.

\begin{figure}
\begin{center}
\includegraphics[width=3.0in]{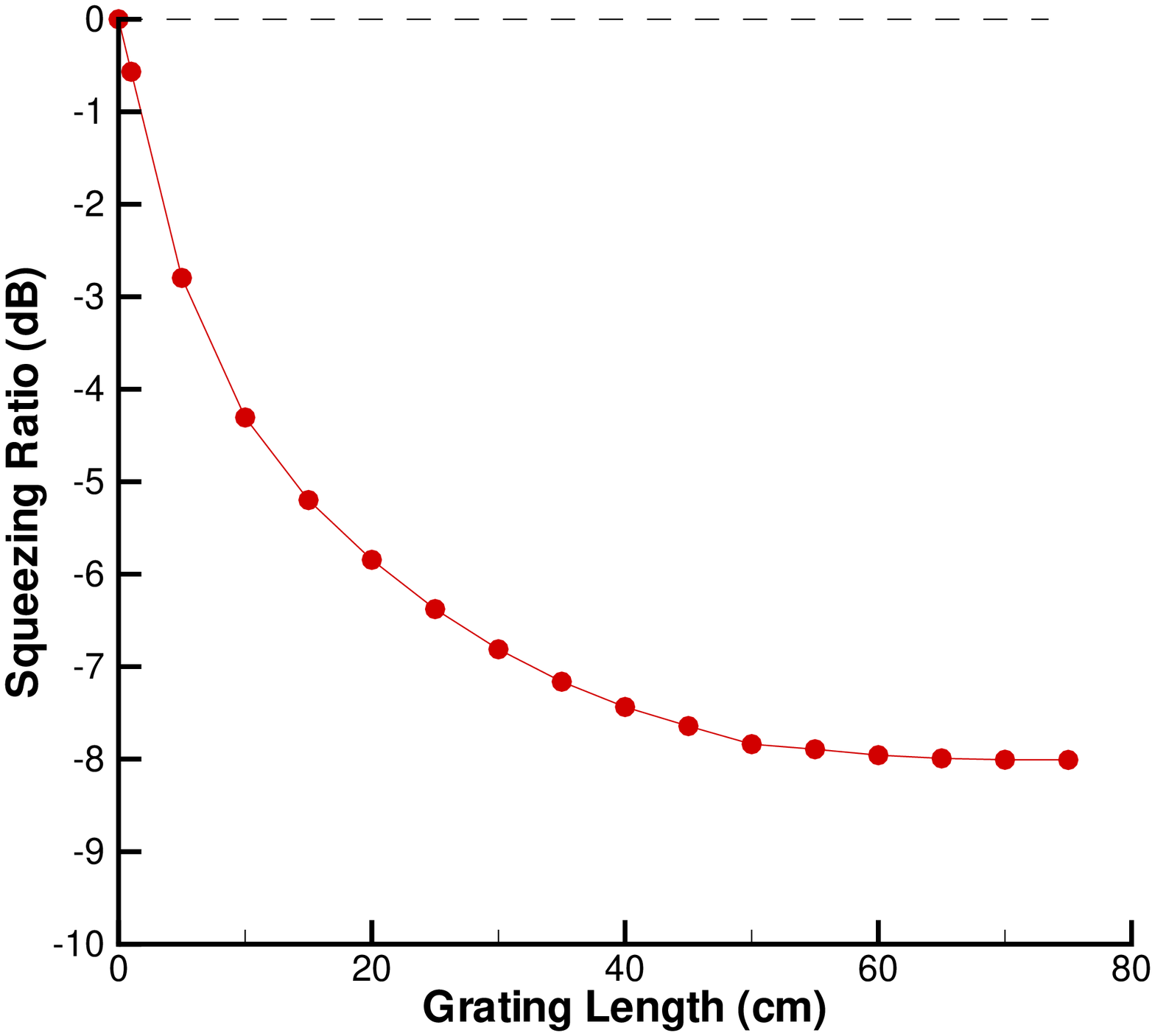} 
\caption{Photon number squeezing ratio for Bragg solitons after propagating through different length of FBGs.}
\label{SRLfig}
\includegraphics[width=3in]{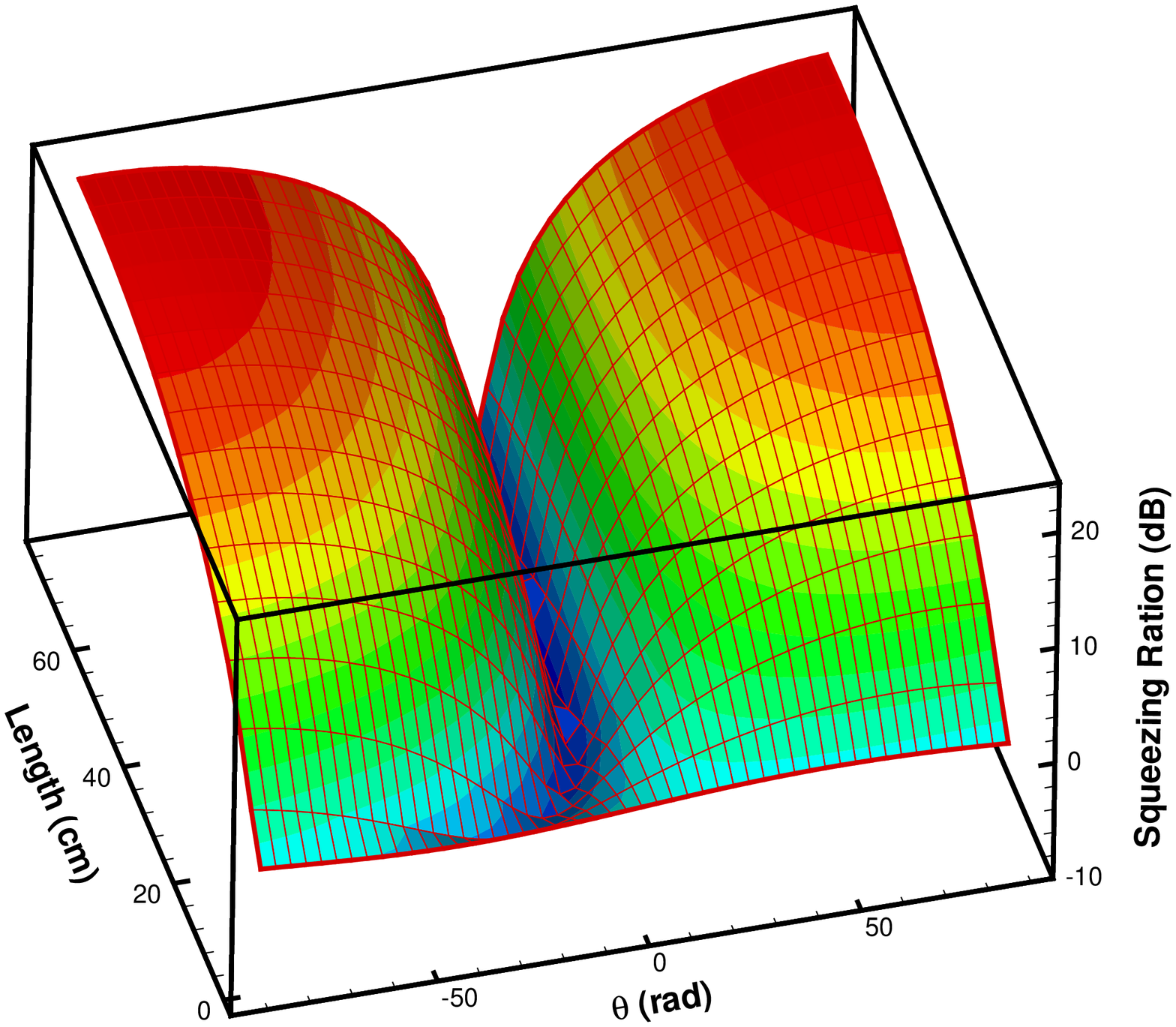}  
\caption{Squeezing ratio of the quadrature field versus the FBG length and the local oscillator phase.}
\label{SSLfig}
\end{center}
\end{figure}

Fig. \ref{SRLfig} shows the dependence of the photon number squeezing ratios for different FBG lengths with a constant input intensity ($I = 4.5 GW/cm^2$).
We find that the squeezing ratio monotonically decreases with the increasing of the FBG length and saturates at the length around 60 $cm$.
Intuitively the filtering effects of the FBG will unavoidably introduce additional noises on the light fields and will eventually cause the squeezing ratio to become saturated.
In Fig. \ref{SSLfig} we also plot the squeezing ratio for different FBG lengths and for different local oscillator phases with a constant input intensity ($I = 4.5 GW/cm^2$).
Again when the FBG length is long enough, the squeezing direction will be in the in-phase (photon number) quadrature.

\section{Quantum FBG solitons in nonuniform gratings}
\label{secAPD}
It is well known that one can engineer the dispersion along the FBG by using a nonuniform FBG.
Such apodized nonlinear FBGs have been used for adiabatic soliton pulse compression within a very short length of several centimeters \cite{Lenz98, Eggleton99}.
Intuitively the solitons with higher peak intensities will exhibit large squeezing due to higher nonlinear effects.
It is thus expected that one should be possible to compress the pulsewidth of the FBG soliton and enhance its squeezing ratio simultaneously with the use of a suitably apodized FBG.
To verify this idea, here we consider a nonuniform FBG which has a position dependent coupling coefficient described by
\begin{eqnarray}
\kappa(z) = \kappa_0 + \alpha z 
\end{eqnarray}
where $\kappa_0 = \omega_0\tilde{\epsilon}/2\bar{n} c$ is the initial coupling coefficient and $\alpha$ is the slope of the coupling coefficient.

\begin{figure}
\begin{center}
\includegraphics[width=3in]{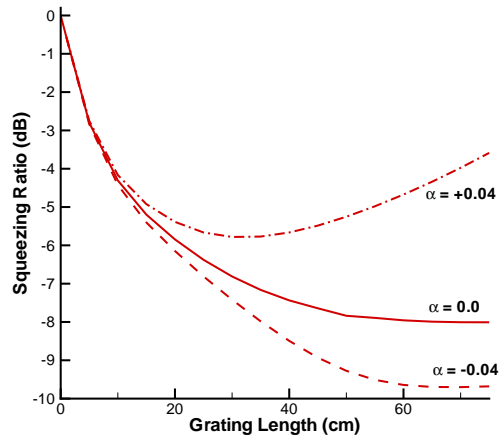}  
\caption{ Photon number squeezing ratio versus the FBG length with a constant input intensity ($I = 4.5 GW/cm^2$). The coupling coefficient of the FBG is a function of propagating length: $\kappa(z) = \kappa_0 + \alpha z$. The three curves correspond to different slopes of $\kappa(z)$; Solid-line: $\alpha = 0$; Dashed-line: $\alpha = -0.04$; Dash-dotted line: $\alpha = + 0.04$.}
\label{SRL-3}
\end{center}
\end{figure}

\begin{figure}
\begin{center}
\includegraphics[width=3in]{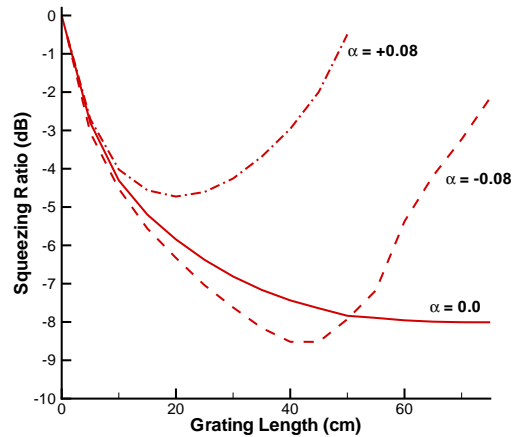} 
\caption{Photon number squeezing ratio versus the FBG length with different slopes of $\kappa(z)$; Solid-line: $\alpha = 0$; Dashed-line: $\alpha = -0.08$; Dash-dotted line: $\alpha = + 0.08$. Other parameters are the same with those in Fig. \ref{SRL-3}}
\label{SRL-5}
\end{center}
\end{figure}

In the following calculation we consider the same 60 $ps$ FWHM {\it sech}-shaped input pulse with the peak intensity of $I = 4.5 GW/cm^2$ for the nonuniform grating without changing any parameter.
In Fig. \ref{SRL-3} we plot the squeezing ratio versus the FBG length with a constant input intensity ($I = 4.5 GW/cm^2$) and different apodization slopes.
The three curves in Fig. \ref{SRL-3} correspond to different slopes of $\kappa(z)$ ($\alpha = +0.04, 0, -0.04$) respectively.
When $\alpha = 0.04 (1/cm^{-2})$, the FWHM of the original FBG soliton (60 $ps$) will be adiabatically compressed down to 30 $ps$ after propagating through a 70 $cm$ FBG.
Because of this, the achievable optimal squeezing ratio thus increases from 8 $dB$ to 9.7 $dB$ for the propagation distance of 70 $cm$.
On the other hand, if the slope of the coupling coefficient is positive, we have a broaden FBG soliton and the optimal squeezing ratio will be degraded.
If the slope of the coupling coefficient is too large, then the soliton cannot be compressed adiabatically and thus the optimal squeezing ratio will be degraded eventually.
This can be seen in Fig. \ref{SRL-5}, where we plot the squeezing ratio versus the grating length with larger slopes ($\alpha = -0.08, +0.08$).
The result for the uniform FBG case ($\alpha = 0$) is also plotted for comparison.
These results show that one can actually tailor the squeezing ratio of the FBG solitons as long as the soliton pulses evolve adiabatically. 
Recently some new theoretical results for the dynamics of classical gap solitons in an apodized grating have been carried out \cite{Mak03}.
It was shown that the velocity of the gap soliton can be essentially slowed down in the apodized grating and interesting soliton collision behaviors including the merge of the solitons were predicted.
We shall investigate more about the other quantum dynamics of FBG solitons in apodized fiber gratings in the future.

\section{Conclusion}
\label{secCon}
In conclusion, by following the line of theoretical development on the quantum theory of optical solitons initiated by Prof. Hermann A. Haus, we have developed a general quantum theory for bi-directional nonlinear optical pulse propagation problems and have especially used it to study the squeezing phenomena of fiber Bragg grating solitons.
It has been shown for the first time that the output FBG soliton pulses will get amplitude squeezed automatically.
The squeezing ratio of the FBG solitons exhibits interesting relation with the fiber grating length as well as with the intensity of the input pulse.
The squeezing ratio saturates after a certain grating length and the optimal squeezing ratio occurs when the intensity of the FBG soliton is slightly large than that of the fundamental soliton.
With the use of nonuniform FBGs, we also find that one can compress the FBG solitons and enhance its squeezing ratio simultaneously, as long as the soliton pulses evolve adiabatically.
To actually measure the quantum fluctuations of the fiber Bragg grating solitons experimentally, we propose to use a time-gating device to block out other smaller multiple transmitted pulses and only directly detect the first transmitted pulse from the grating.
Since our calculation is already based on the existing experimental parameters, it shall be very interesting to see if one can actually verify the theoretical predictions experimentally.

\section*{Acknowledgments}
The authors want to use this paper to express their deep thanks and respects to Prof. Hermann A. Haus for the inspirations we have directly or indirectly learned from him.

\end{document}